\documentclass[epj]{webofc}
\usepackage[varg]{txfonts}   
\woctitle{International Nuclear Physics Conference INPC2013: 
2--7 June, 2013, Firenze, Italy}
\begin{document}
\title{Production of Energetic Light Fragments in Spallation Reactions}
%
%

\author{Stepan G. Mashnik\inst{1}\fnsep\thanks{\email{mashnik@lanl.gov}} \and
        Leslie M. Kerby\inst{1,2}
\and
         Konstantin K. Gudima\inst{3}
\and
         Arnold J. Sierk\inst{1}
}

\institute{Los Alamos National Laboratory, Los Alamos, NM 87545, USA 
\and
           University of Idaho, Moscow, Idaho 83844-4264, USA 
\and
           Institute of Applied Physics,
Academy of Science of Moldova, Chi\c{s}in\u{a}u, Moldova
          }

\abstract{%
Different reaction mechanisms contribute to the production
of light fragments (LF) from nuclear reactions.
Available models cannot accurately predict
emission of LF from arbitrary reactions.
However, the emission of LF
is important for many applications, such as
cosmic-ray-induced single event upsets, radiation protection, and
cancer therapy with proton and heavy-ion beams, to name just a few.
The cascade-exciton model (CEM) and the Los Alamos version
of the quark-gluon string model (LAQGSM), as implemented in the
CEM03.03 and LAQGSM03.03 event generators used in
the Los Alamos Monte Carlo transport code MCNP6,
describe quite well the spectra of fragments with sizes up to $^{4}$He 
across a broad range of target masses and incident energies. 
However, 
they do not predict high-energy tails for LF heavier than $^4$He. 
The 
standard
versions of CEM and LAQGSM do
not account for preequilibrium emission of LF larger than $^{4}$He.
The aim of our work is to extend the preequilibrium 
model to include such processes.
We do this
by including the emission of fragments heavier than $^4$He at the 
preequilibrium stage, and using
an improved version of the Fermi Break-up model,
providing improved agreement with various experimental data.  
}
\maketitle
\section{Introduction}
\label{intro}

Emission of light fragments (LF) from nuclear reactions is an 
interesting
open question. 
Different reaction mechanisms contribute to their production; the relative 
roles of each, and how they change with incident energy, mass number of 
the target, and the type and emission energy of the fragments are not 
completely understood.

The cascade-exciton model (CEM) \cite{CEM, Trieste08} 
version 03.03 and the Los Alamos version of the quark-gluon string 
model (LAQGSM) \cite{Trieste08, LAQGSM} version 03.03 event 
generators in the Los Alamos transport code 
MCNP6 \cite{MCNP6} describe quite well the spectra of fragments with 
sizes up to $^{4}$He across a broad range of target masses and incident 
energies (up to $\sim 5$~GeV for CEM and up to $\sim 1$~TeV/A for LAQGSM). 
However, they do not predict the high-energy tails of LF spectra heavier 
than $^4$He well. Most LF with energies above several tens of MeV are 
emitted during the precompound stage of a reaction. The 
standard
versions 
of the CEM and LAQGSM event generators do not account for precompound 
emission of these heavier LF. 

The aim of our work is to extend the precompound model to include 
such processes, leading to an increase of predictive power of LF-production 
in MCNP6. This entails upgrading the modified exciton model (MEM) currently used 
at the preequilibrium stage in CEM and LAQGSM. It will also include expansion 
and examination of the coalescence and Fermi break-up models used in the 
precompound stages of spallation reactions within CEM and LAQGSM.  
Extending our models in this way has 
provided preliminary results that have much
better agreement with experimental data.

\section{Theoretical Background}
\label{sec-1}
These models consider that
a reaction begins with the intranuclear cascade, referred to 
as the INC.
The incident particle or nucleus 
(in the case of LAQGSM) enters the target nucleus and begins 
interacting with nucleons, scattering off them and also often creating 
new particles in the process. The incident particle and all newly 
created particles are followed until they either escape from the 
nucleus or reach a threshold energy 
and are then considered ``absorbed" by the nucleus. 

The preequilibrium stage uses the modified exciton model (MEM) to 
determine emission of protons, neutrons, and fragments up to $^4$He 
from the residual nucleus. 
In the evaporation stage nucleons in the outer shells of 
the residual nucleus can ``evaporate" off, either singly or as fragments. 
The CEM evaporation stage is modeled with
a modification of Furihata's generalized 
evaporation model code (GEM2) \cite{GEM2}, and can emit light fragments 
up to $^{28}$Mg.
During and after evaporation, the code looks to see if there is a 
nuclide with $Z \geq 65$ which is thus fissionable. If this nuclide 
is randomly determined to fission, the code allows for evaporation from
the fission fragments.

There are two models that are not directly part of this linear 
progression just outlined: coalescence and Fermi break-up. The INC 
stage only emits nucleons and pions (and other particles, 
in the case of LAQGSM at high energies), so the coalescence 
model ``coalesces" some of the nucleons produced in 
the INC into larger fragments by comparing their momenta. If their 
momenta are similar enough then they coalesce. The current coalescence 
model can only coalesce up to a $^4$He fragment, the same as the 
preequilibrium stage. The Fermi break-up is a very simplified 
multifragmentation model that is fast and accurate for small atomic 
numbers; in the standard CEM and LAQGSM models it is used when any 
nuclide has a mass number less than 13.

More details on the models can be found in Refs. \cite{CEM, Trieste08, LAQGSM}.
As the MEM uses a Monte-Carlo technique to solve the master equations describing
the behavior of the nucleus at the preequilibrium stage (see details in 
\cite{CEM}), it is very easy to extend the number of types of possible
LF that can be emitted. We have extended the MEM to consider
emission of up to 66 types of nucleons and LF, up to $^{28}$Mg
\cite{Kerby2012}.
As a starting point, for the inverse cross sections, Coulomb barriers, and
binding energies
of all LF we use the approximations adopted by GEM2 \cite{GEM2}.

\section{Results and Conclusion}
\label{sec-2}
Expanding the Fermi break-up model to include heavier LF (up to $A = 16$) 
yields increased accuracy for reactions with lighter targets. Figs. 1 and 2
provide examples 
of calculations by our updated CEM and LAQGSM models, respectively
compared to experimental data \cite{Uozumi07}--\cite{Lemaire78}.
As can be seen, results from the expanded model achieve good 
agreement with experimental results for these light nuclei.

Expanding the MEM to include heavier LF (up to $^{28}$Mg) yields 
increased accuracy for reactions on medium and heavy nuclei. 
Figs. 3 and 4 
compare our simulations using the expanded MEM with data by 
Green et al. \cite{Green87} and Budzanowski et al. \cite{Budzanowski10}.

Similar results for different LF spectra are obtained for several
other reactions (see, e.g., \cite{Kerby2012}).

\begin{figure}
\centering
\sidecaption
\includegraphics[width=9.7cm,clip]{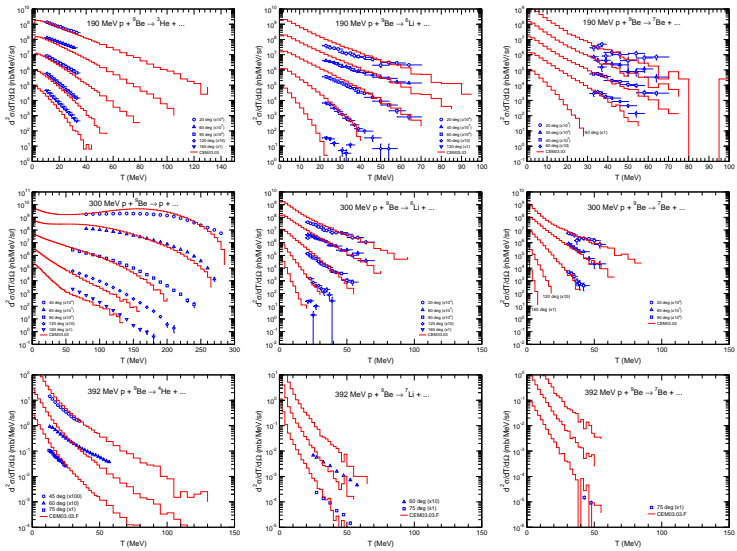}
\caption{Examples of measured particle and LF double-differential
spectra from p+$^{9}$Be at 190, 300, and 392 MeV \cite{Uozumi07} (symbols)
compared with our CEM results (histograms). All the LF from these reactions
are calculated by CEM either using the Fermi break-up model after the INC
or as final products (residual nuclei) after the INC and Fermi break-up
stages of interactions (Fermi break-up is used for nuclei with $A <13$
instead of using preequilibrium emission and/or evaporation of particles).}
\label{fig-1}       
\end{figure}

\begin{figure}
\centering
\sidecaption
\includegraphics[width=6.6cm,clip]{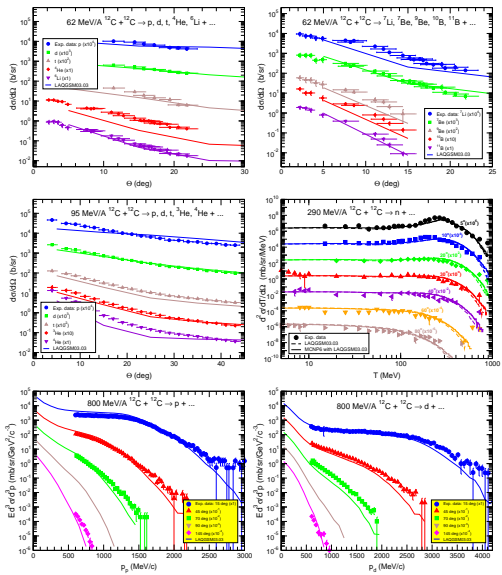}
\caption{Examples of measured particle and LF angular, double-differential,
and invariant spectra from $^{12}$C+$^{12}$C at 62 MeV/A \cite{DeNapoli12},
95 MeV/A \cite{Dudouet13, Dudouet13a}, 290 MeV/A \cite{Iwata01}, 
and 800 MeV/A \cite{Lemaire78} (symbols)
compared with LAQGSM03.03 results and calculations by MCNP6 using LAQGSM03.03.
All the LF from these reactions
are calculated by LAQGSM either using the Fermi break-up model after the INC
or as final products (residual nuclei) after the INC and Fermi break-up
stages of the interactions (Fermi break-up is used for nuclei with $A <13$
instead of using preequilibrium emission and/or evaporation of particles). 
Note that our INC does not account for $\alpha$-clustering
in $^{12}$C and is not well grounded at low energies, therefore the calculated
spectra of $^4$He are under-predicted and generally, the higher the incident
energy, the better the agreement with the experimental data.}
\label{fig-2}       
\end{figure}

\begin{figure}
\centering
\sidecaption
\includegraphics[width=6.6cm,clip]{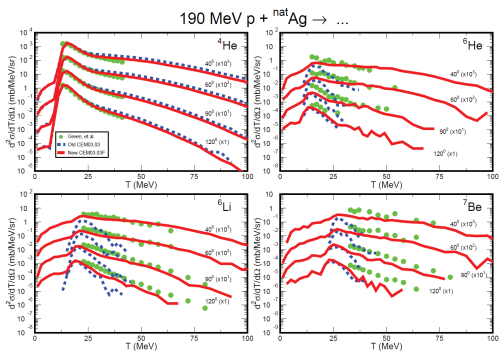}
\caption{Examples of measured LF double-differential
spectra from 190 MeV p + Ag \cite{Green87} (green circles)
compared with results by the extended CEM (solid red lines)
and by the older ``standard'' CEM (blue dashed lines). All the energetic 
LF heavier than $^4$He
from these reactions
are calculated by CEM 
with the extended preequilibrium model.
Similar results are obtained
for other LF spectra
from this reaction as well as for the
p + Ag reactions measured 
by Green et al. at 
300 and 480 MeV \cite{Green87, Green84}.}
\label{fig-3}       
\end{figure}

\begin{figure}
\centering
\sidecaption
\includegraphics[width=9.5cm,clip]{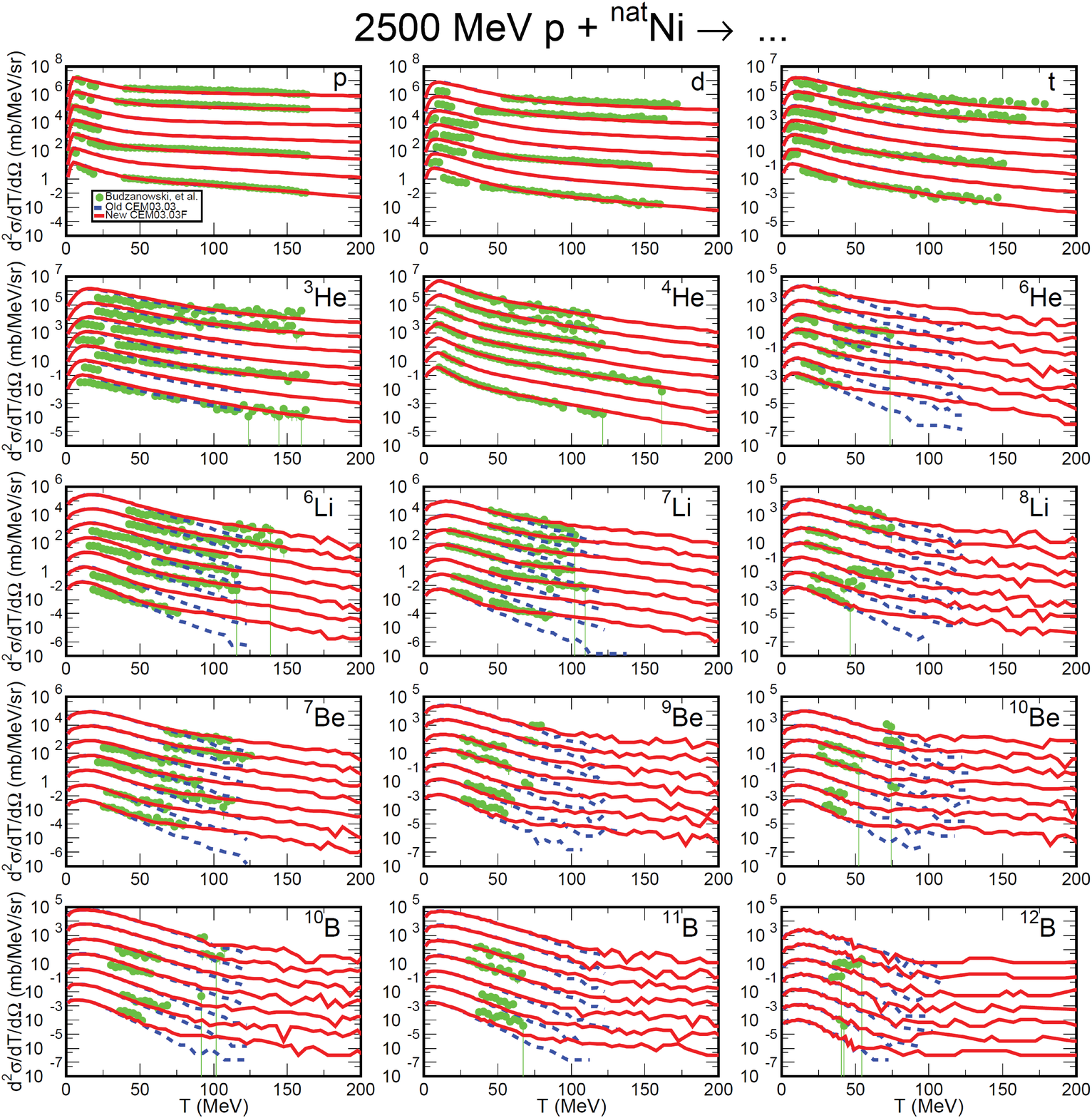}
\caption{Measured LF double-differential
spectra from 2.5 GeV p + Ni \cite{Budzanowski10} (green circles)
at 15.6, 20, 35, 50, 65, 80, and 100 degrees
(multiplied, respectively, by 10$^6$, 10$^5$, 10$^4$, 10$^3$,
10$^2$, 10$^1$, and 1)
compared 
with results by the extended CEM (solid red lines)
and by the older ``standard'' CEM (blue dashed lines). Note that all energetic 
LF heavier than $^4$He
from these reactions
are calculated by CEM 
with the extended preequilibrium model.
Similar results are obtained
for several other LF spectra
from this reaction as well as for the
p + Ni reactions measured 
by Budzanowski et al. at 
1.2 and 1.9 GeV \cite{Budzanowski10}.}
\label{fig-4}       
\end{figure}

Our results indicate that expanding the MEM used by the CEM and LAQGSM
event generators of MCNP6
 to include LF preequilibrium 
emission significantly increases accuracy of the high-energy spectra compared 
to experimental data.

We thank Jeremie Dudouet for providing prepublication numerical values of the
measurements presented in Ref. \cite{Dudouet13a}.
This study was carried out under the auspices of the National Nuclear Security 
Administration of the U.S. Department of Energy at Los Alamos National Laboratory 
under Contract No. DE-AC52-06NA253996.

\end{document}